\documentclass[aps,prl,twocolumn,superscriptaddress,longbibliography]{revtex4-2}
\usepackage[colorlinks=true, citecolor=blue, urlcolor=blue, linkcolor=red]{hyperref}
\usepackage{amssymb,amsmath,orcidlink,physics}
\renewcommand{\section}[1]{{\par\it #1.---}\ignorespaces}

\begin{document}
\title{Hybrid-order nonlinear topological phases}
\author{Yu-Peng Ma\orcidlink{0009-0006-9911-9424}}
\affiliation{Key Laboratory of Quantum Theory and Applications of Ministry of Education, Lanzhou Center for Theoretical Physics, Gansu Provincial Research Center for Basic Disciplines of Quantum Physics, Key Laboratory of Theoretical Physics of Gansu Province, Lanzhou University, Lanzhou 730000, China}
\author{Ming-Jian Gao\orcidlink{0000-0002-6128-8381}}
\affiliation{Key Laboratory of Quantum Theory and Applications of Ministry of Education, Lanzhou Center for Theoretical Physics, Gansu Provincial Research Center for Basic Disciplines of Quantum Physics, Key Laboratory of Theoretical Physics of Gansu Province, Lanzhou University, Lanzhou 730000, China}
\author{Jun-Hong An\orcidlink{0000-0002-3475-0729}}\email{anjhong@lzu.edu.cn}
\affiliation{Key Laboratory of Quantum Theory and Applications of Ministry of Education, Lanzhou Center for Theoretical Physics, Gansu Provincial Research Center for Basic Disciplines of Quantum Physics, Key Laboratory of Theoretical Physics of Gansu Province, Lanzhou University, Lanzhou 730000, China}
\begin{abstract}
The bulk-boundary correspondence (BBC), which relates topological invariants to boundary modes, is well understood for linear systems but remains an open question in the presence of nonlinearity, where multigap topologies make the BBC obscure and the topological description troublesome. We address this by developing an auxiliary-system formalism that enables topological classification of nonlinear-eigenvalue systems. In the two-dimensional (2D) case, we show that two fixed eigenvalues can harbor first-order gapless boundary modes and second-order corner modes. Stacking this 2D system along the third dimension (3D) reveals distinct hybrid-order realizations. Under uniform stacking, the corner states in the $xy$ plane transform into hinge states along the $z$ axis, yielding a 3D second-order phase, while the gapless boundary states become side-surface states, yielding a 3D first-order phase. For dimerized stacking, these states are further confined to the two ends of the $z$ axis, yielding a 3D third-order phase, and localized at the hinges, yielding a 3D second-order phase. Our results establish a multigap bulk-boundary correspondence and identify stacking engineering as a versatile platform for exploring hybrid-order topological phases in nonlinear systems.
\end{abstract}
\maketitle

\section{Introduction}
The discovery and development of topological phases have opened a new chapter in condensed matter physics \cite{RevModPhys.82.3045,RevModPhys.83.1057,RevModPhys.88.021004,RevModPhys.88.035005,RevModPhys.89.040502,PhysRevX.7.041069}. Unlike Landau phase transitions, topological phase transitions do not rely on symmetry breaking \cite{PhysRevLett.95.226801,PhysRevLett.95.146802}. They all have boundary states within the band gap that are protected by symmetries and robust to disorder. The relation between the bulk and the boundary is known as the bulk-boundary correspondence (BBC). It indicates that these boundary states are essentially guaranteed by the topology of the bulk bands  \cite{PhysRevLett.132.136401,Zhu2025,Li2024,Hossain2024,PhysRevX.14.041048}.  Within this standard band-theory framework, the BBC can allow different types of topological boundary states to appear simultaneously in the same physical system. Examples include one-dimensional gapless boundary states and corner states in hybrid-order topological systems \cite{PhysRevLett.126.156801,l1n5-1jsm,PhysRevA.98.043838,PhysRevB.110.155415,Li2023}. The coexistence is important because it combines the functions of different orders of boundary states in one system, which is beneficial for the application design of topological devices. The hybrid-order topological phases has been realized or proposed in electronic materials \cite{Zhang2025,doi:10.1126/science.adk1270,doi:10.1126/science.adj3742,Bouhon2020}, photonic systems \cite{Lu2014,Huang2024,doi:10.1126/science.adr5234,33mm-mx88,PhysRevA.98.043838}, ultracold atoms \cite{Jotzu2014,PhysRevLett.131.263001,Braun2024}, circuit systems \cite{PhysRevX.5.021031,Lenggenhager2022,Wang2020,Lee2018}, and acoustic systems \cite{Kane2014,Xue2019,Cheng2025,PhysRevLett.132.216602,PhysRevApplied.18.034066,PhysRevLett.128.224301,Xue2020,PhysRevLett.129.254301}.

Recently, a great deal of attention has turned from linear to nonlinear topological systems, which have two classes. The first class consists of nonlinear eigenstate problems in which the eigen equation depends nonlinearly on the eigenstate amplitude. This type of nonlinearity appears in Kerr nonlinear photonic systems \cite{PhysRevLett.117.143901,Sone2024,Szameit2024,PhysRevLett.134.093801}, topolectric circuits with nonlinear electronic components \cite{Hadad2018,PhysRevResearch.5.L012041}, and mechanical metamaterials based on nonlinear springs \cite{PhysRevB.100.014302}. Nonlinearity in topological boundary modes gives rise to distinctive phenomena, including soliton physics \cite{PhysRevLett.100.013905,Zhang2020,PhysRevX.11.041057,PhysRevLett.128.093901,PhysRevB.107.184313,Choi2024,coen2024nonlinear,Hashemi2024,Mittal2021}, synchronization \cite{PhysRevResearch.4.023211,Moille2025,https://doi.org/10.1002/advs.202408460}, and potential topological-laser implementations \cite{Leefmans2024}. The second class is nonlinear-eigenvalue systems, where nonlinearity is present via the eigenvalue. This type of nonlinearity appears naturally in dispersive photonic, phononic, and circuit systems  \cite{PhysRevB.50.16835,PhysRevA.78.033834,8937095,8358010,PhysRevE.106.035304,LI2022108835,photonics10050523,PhysRevA.109.043518,PhysRevB.110.174305,PhysRevA.94.022105,PhysRevB.111.045137,DEMESY2020107509,Pocock2018}. The nonlinear nature of the eigenvalues makes the BBC unclear, so the conventional topological framework breaks down. Recently, a method was proposed to reveal the BBC of the real-band topological phases of the nonlinear-eigenvalue system by mapping the system to an auxiliary linear system \cite{Güttel_Tisseur_2017,10.1145/2427023.2427024,PhysRevLett.132.126601,PhysRevB.109.134201}. Within this framework, a BBC beyond conventional band theory has been established for nonlinear-eigenvalue problems \cite{PhysRevLett.132.126601,PhysRevB.109.134201,PhysRevB.111.064310}. However, existing studies mainly describe the nonlinear topological phases with a single-gap boundary modes. A complete description of the multiple-gap topologies in nonlinear-eigenvalue systems is still absent.     

Here, we address the issue of whether different orders of topological boundary modes emerge in multiple gaps of one single nonlinear-eigenvalue system or not. Via investigating a two-dimensional (2D) spinless square-lattice model, we find a class of hybrid-order topological phases in nonlinear-eigenvalue systems. A complete topological characterization is established by the introduction of an auxiliary system. In the 2D case, we find the simultaneous presence of gapless boundary states and gapped corner states within distinct band gaps of the same nonlinear eigenvalue spectrum. After stacking the 2D system into a three-dimensional (3D) one, we uncover distinct 3D hybrid-order topological phases depending on the stacking configuration. In the dimerized-stacked configuration, the corner states in the $xy$ plane become further localized at the two ends of the $z$ axis, manifesting as corner states of a 3D third-order topological phase, while the gapless boundary states in the $xy$ plane localize at the hinges of the $z$ axis, manifesting as a 3D second-order topological phase. In the uniform-stacked case, the boundary states evolve into side-surface states and the corner states transform into hinge states, indicating a coexisting first- and second-order topological phase. Our results demonstrate a multigap bulk-boundary correspondence and uncover hybrid-order topological phases in nonlinear-eigenvalue systems.

\begin{figure}[ttp]
\begin{center}
\includegraphics[width=\columnwidth]{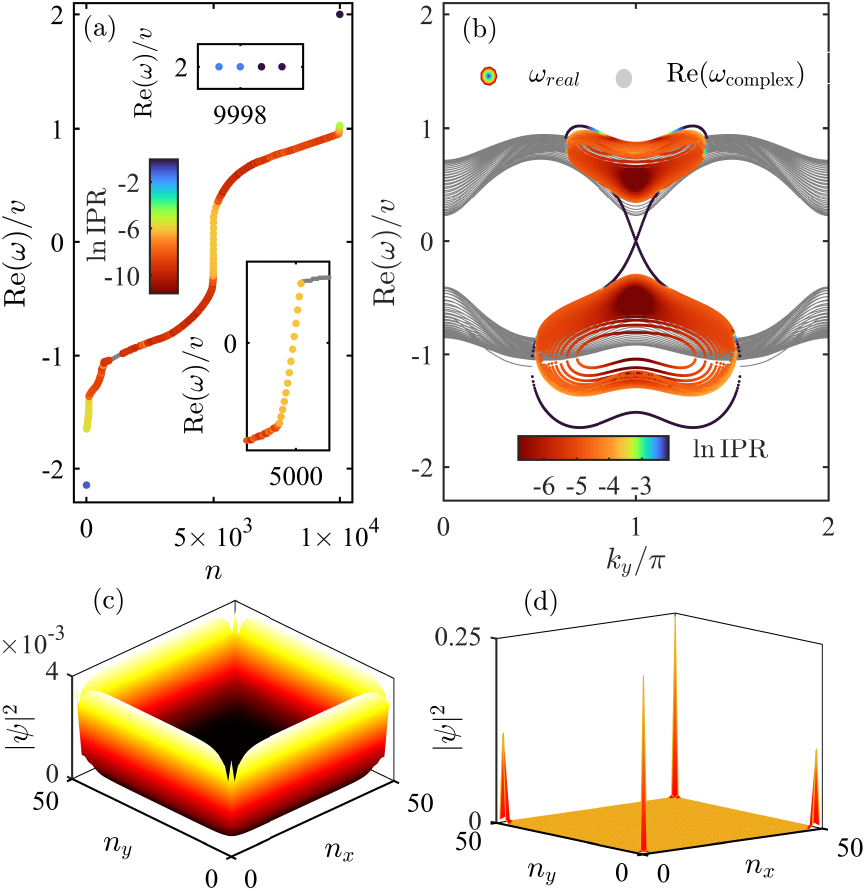}
\end{center}
\caption{Hybrid-order topological phase in a nonlinear-eigenvalue system.
Eigenvalue spectra (a) under the OBC to show the coexisting first-order boundary states at $\omega = 0$ and second-order corner states at $\omega = -M^{-1}$ and (b) under the OBC along the $x$-axis and the PBC along the $y$-axis to confirm the presence of double-degenerate boundary modes at $\omega = 0$. The colored points denote the IPR of the eigenstates with real eigenvalues. The gray points denote the real part of complex eigenvalues. Probability distributions of (c) the first-order boundary state at $\omega = 0$ and (d) the second-order corner state at $\omega=-M^{-1}$. We use $M=-0.5v^{-1}$, $w=0.5v$, $m=1.5v$, $r=v$, and $n_x=n_y=50$.} \label{F1}
\end{figure}

\section{Nonlinear-eigenvalue topological phases}
We consider a class of topological systems whose eigen equation reads
\begin{equation}
H|\psi \rangle=\omega\mathcal{S}(\omega)|\psi \rangle,\label{NH}
\end{equation}
where $H$ is a Hermitian Hamiltonian, $\mathcal{S}(\omega)$ is called the overlap matrix, and $\omega$ is the eigenvalue. The $\omega$-dependence of $\mathcal{S}(\omega)$ renders Eq.~\eqref{NH} nonlinear \cite{Ikramov1993,PhysRevLett.132.126601}. It arises naturally in photonic systems, where the dependence of permittivity on the frequency of propagating waves results in the presence of $\mathcal{S}(\omega)$ \cite{PhysRevB.50.16835,PhysRevA.78.033834,8937095,8358010,PhysRevE.106.035304,LI2022108835,photonics10050523,PhysRevA.109.043518,HOENDERS2025229}. Due to the nonlinearity, some of the eigenvalues $\omega$ may become complex even though $H$ and $\mathcal{S}(\omega)$ are Hermitian \cite{52wp-42b4}. This causes that one cannot use the well-developed Bloch band theory in linear systems to construct the BBC of the topological phases in nonlinear-eigenvalue systems. In order to establish the BBC, we introduce an auxiliary system defined by
\begin{equation}
P(\omega)|\phi\rangle=\lambda|\phi\rangle, \label{PH}
\end{equation}
where $P(\omega)=H-\omega\mathcal{S}(\omega)$ is called the pencil matrix \cite{Ikramov1993} and $\lambda$ is its eigenvalue \cite{PhysRevLett.132.126601,PhysRevA.78.033834,PhysRevA.111.042201}. Equation \eqref{PH} recovers Eq.~\eqref{NH} under the condition $\lambda=0$. Thus, we succeed in converting the nonlinear-eigenvalue
problem into a formally linear one. Such a conversion is exact. Treating $\omega$ as a free parameter and running it in a physically permitted regime for given system-parameter values, we solve Eq. \eqref{PH} and obtain the $\lambda$ spectrum of the auxiliary system. Matching the $\omega$ spectrum, the zero line of the $\lambda$ spectrum shares the same topological feature as the original system. When the $\lambda$ bands are topological, they possess gapless boundary states under
the open-boundary condition (OBC). These states cross $\lambda=0$ so that they also emerge in the original system \cite{PhysRevLett.132.126601}. Therefore, we can define the topological invariant in the auxiliary system to characterize the boundary states and reveal the BBC of the original system.

\section{2D hybrid-order nonlinear topological phases}
We explicitly study a nonlinear spinless system on a square lattice with two degrees of freedom per unit cell. Its Hamiltonian reads
\begin{eqnarray}
H=\sum_{\mathbf r}\big[mc_{\mathbf r}^\dagger \sigma_z c_{\mathbf r}
+w\big( c_{\mathbf r+\hat x}^\dagger T_+ c_{\mathbf r} +c_{\mathbf r+\hat y}^\dagger T_- c_{\mathbf r} +\text{h.c.} \big)\big],  \label{HM}
\end{eqnarray}
where $T_{\pm}=\sigma_z-i\sigma_x\pm i\sigma_y$, ${\mathcal S}(\omega)=\sigma_0+\omega s_1$, and $c_{\mathbf r}=(c_{{\mathbf r},A},c_{{\mathbf r},B})^T$. The nonlinear term is $s_1=\sum_{\mathbf r} \{c_{\mathbf r}^\dagger\big(M\sigma_0+v\sigma_z/4\big)c_{\mathbf r}+r[(c_{\mathbf r+2\hat x}^\dagger +c_{\mathbf r+2\hat y}^\dagger) \tilde{\sigma}_- c_{\mathbf r}-ic_{\mathbf r+\hat x-\hat y}^\dagger \sigma_{y} c_{\mathbf r}+\text{h.c.}]\}$, where $\tilde{\sigma}_-=(i\sigma_x-\sigma_z)/2$ \cite{PhysRevLett.123.177001,PhysRevLett.132.126601,PhysRevB.107.235132}. Here, $w$ and $r$ are the linear and nonlinear hopping amplitudes, $m$ and $v$ are the linear and nonlinear masses, $M$ is the nonlinear strength, $\sigma_{x,y,z}$ are the Pauli matrices, and $\sigma_0$ is the identity matrix. Equation \eqref{NH} is rewritten as
\begin{equation}
\left( \begin{array}{cc}0 & I  \\ s_{1}^{-1}H & -s_{1}^{-1}\sigma_0 \end{array}\right)\left( \begin{array}{c}|\psi\rangle \\ 
\omega |\psi\rangle \end{array}\right)=\omega\left( \begin{array}{c}|\psi\rangle   \\ \omega |\psi\rangle \end{array}
\right).\label{linear}
\end{equation} 
The presence of $s_1$ turns the system into a four-band model. The $\omega$ spectrum is obtained by numerically solving Eq. \eqref{linear}. Different from the previous nonlinear topological phases \cite{Sone2024}, whose topological invariant can only be defined in the real-space eigenstates, we can develop the momentum-space topological description of our nonlinear topological phases. To reveal the BBC, we examine the pencil matrix under the periodic-boundary condition (PBC), which reads $P(\mathbf{k},\omega)=\sum_{j=0,\pm,z}p_j(\omega)\sigma_j$,
with $\sigma_\pm=\big(\sigma_x\pm i\sigma_y\big)/2$,
$p_0(\omega)=-\big(\omega+M\omega^2\big)$, $p_\pm(\mathbf{k},\omega)=2w\big[\big(1\pm i\big)\sin k_x+\big(1\mp i\big)\sin k_y\big]+\omega^2 r\big[\sin 2k_x+\sin 2k_y \pm 2i\sin(k_x-k_y)\big]$, and $p_z(\mathbf{k},\omega)=m+2 w \big(\cos k_x+\cos k_y\big)-\omega^2 \big[v/4-r\big(\cos 2k_x+\cos 2k_y\big)\big]$. 
At the high-symmetry points of $k_x$ and $k_y$ being zero or $\pi$, the bands of $P({\bf k},\omega)$ touch at $p_0(\omega)$ when $m=\omega^2(v/4-2r)$ and $\omega^2(v/4-2r)\pm 4w$. Another band touching at $p_0(\omega)$ occurs when $m=\omega^2(v/4+2r)$ under $|w/(\omega^2r)|\leq 1$. Because only the zero line $p_0(\omega)=0$ gives the $\omega$ spectrum, the modes of the original system hosting the topological phases are $\omega_\star=0$ and $-M^{-1}$. Thus, the conditions for the topological phase transition are recast into
\begin{eqnarray}
    m&=&\omega_\star^2(v/4-2r),\label{dshl}\\
    m&=&\omega_\star^2(v/4-2r)\pm 4w,\label{sld}\\
    m&=&\omega_\star^2(v/4+2r)~\text{when}~|w/(\omega_\star^2r)|\leq 1.\label{ddsd}
\end{eqnarray}
At the zero mode, the components of the Bloch vector of $P(\mathbf{k},0)$ become $p_\pm(\mathbf{k},0)=2w\big[\big(1\pm i\big)\sin k_x+\big(1\mp i\big)\sin k_y\big]$ and $p_z(\mathbf{k},0)=m+2 w \big(\cos k_x+\cos k_y\big)$. It is a Chern insulator characterized by the Chern number 
\begin{equation}
    {\mathcal C}_1=\int_{\rm BZ} \frac{{\rm d}^2{\bf k}}{2\pi}\big[i\langle\partial_{k_x}\phi({\bf k},0)|\partial_{k_y}\phi({\bf k},0)\rangle+\text{c.c.}\big],
\end{equation} where $|\phi({\bf k},0)\rangle$ is the eigenstate of $P({\bf k},0)$. The components of the Bloch vector of pencil matrix for the $\omega_\star=-M^{-1}$ mode are $p_\pm(\mathbf{k},-M^{-1})=2w\big[\big(1\pm i\big)\sin k_x+\big(1\mp i\big)\sin k_y\big]+\frac{r}{M^2}\big[\sin(2k_x)+\sin(2k_y)\pm 2i\sin(k_x-k_y)\big]$, and $p_z(\mathbf{k},-M^{-1})=m+2 w \big(\cos k_x+\cos k_y\big)-\frac{1}{M^2} \big[v/4-r\big(\cos 2k_x+\cos 2k_y\big)\big]$. Having the same form as the Hamiltonian of the second-order topological superconductor~\cite{PhysRevLett.123.177001}, $P(\mathbf k,-M^{-1})$ supports a higher-order topological phase. It satisfies the particle-hole symmetry $\Xi P(\mathbf k,-M^{-1})\Xi^{-1}=-P(-\mathbf k,-M^{-1})$ with $\Xi=\sigma_x\mathcal K$ and mirror-rotation symmetry $U_{xy}P(k_x,k_y,-M^{-1})U_{xy}^{-1}=-P(k_y,k_x,-M^{-1})$ with $U_{xy}=\sigma_y$. Thus, its second-order topology is characterized by the mirror-graded winding number defined along the high-symmetry line $k_x=k_y\equiv k$ as
\begin{equation}
    {\mathcal W_2}=\int_{-\pi}^{\pi} \frac{{\rm d}k}{2\pi i}\langle \phi(k,-M^{-1})|\partial_{k}\phi(k,-M^{-1})\rangle,
\end{equation}where $|\phi(k,-M^{-1})\rangle$ is the eigenstate of $P(k,k,-M^{-1})$. Its first-order topology is also described by the Chern number ${\mathcal C}_2$.

\begin{figure}[ttp]
\begin{center}
\includegraphics[width=\columnwidth]{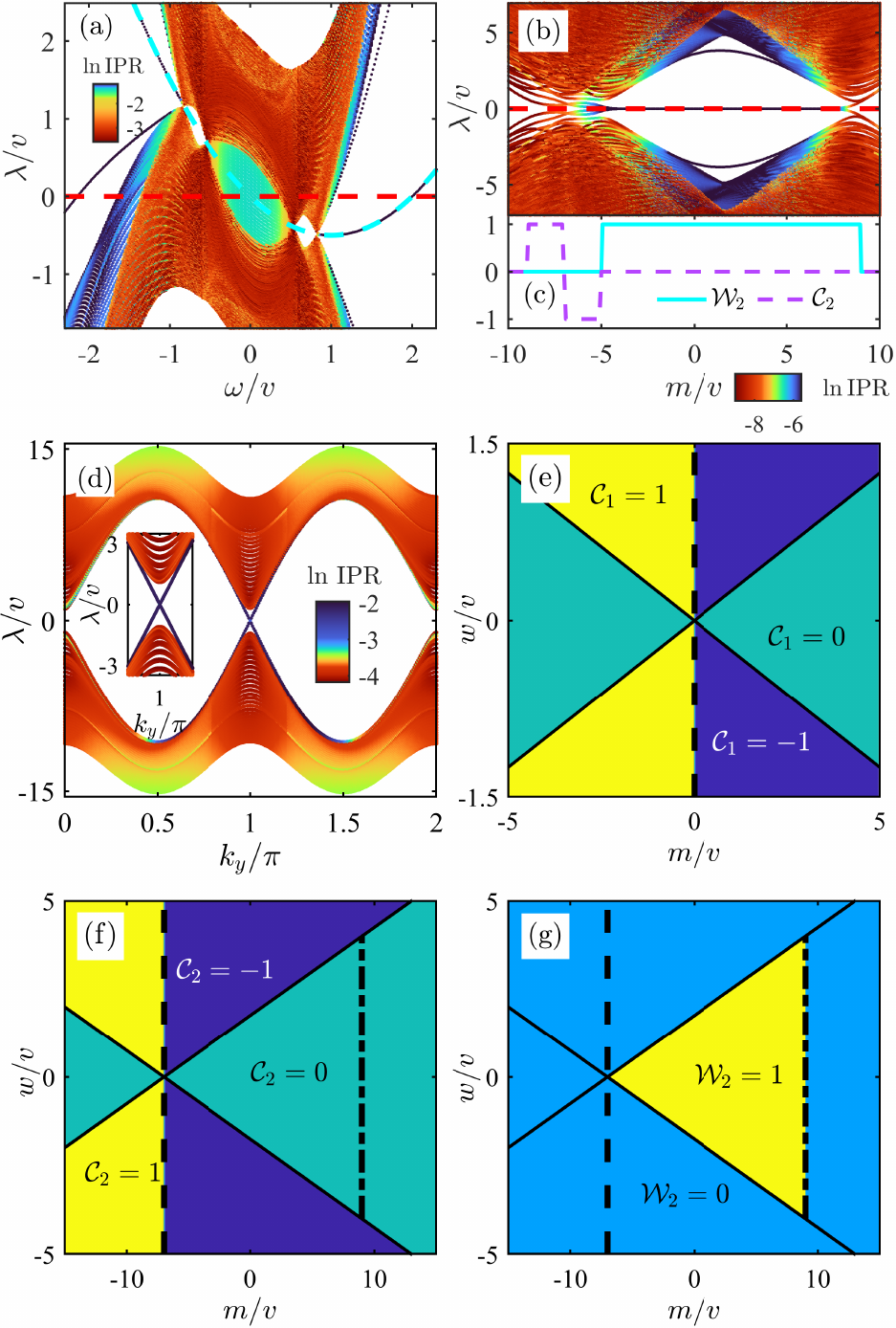}
\end{center}
\caption{Spectrum and topological invariants of the auxiliary system.
(a) Eigenvalues of $P(\mathbf{k},\omega)$ in different $\omega$ when $m=1.5v$. The red dashed line indicates $\lambda=0$, whose intersections with the spectrum reproduce the real-$\omega$ spectrum in Fig.~\ref{F1}(a). The cyan dashed line denotes $p_0(\omega)$, which equals to zero just at $\omega_\star=0$ and $-M^{-1}$. (b) Eigenvalues and (c) Chern number $\mathcal{C}_2$ and mirror-graded winding number $\mathcal{W}_2$ of $P(\mathbf{k},-M^{-1})$ in different $m$. (d) Eigenvalues of the auxiliary system in the strip geometry when $m=-6v$ to confirm the presence of first-order boundary states at $\lambda = 0$. Diagrams of the first-order topological phases characterized by (e) ${\cal C}_1$ at $\omega_\star=0$ and (f) $\mathcal{C}_2$ at $\omega_\star=-M^{-1}$, and (g) the second-order topological phases characterized by $\mathcal{W}_2$ at $\omega_\star=-M^{-1}$. In (e)-(g), the dashed, solid, and dot-dashed lines denote Eqs. \eqref{dshl}, \eqref{sld}, and \eqref{ddsd}, respectively. Other parameters are the same as Fig. \ref{F1}.}  \label{F2}
\end{figure}

Figure~\ref{F1}(a) shows the real $\omega$ and the real part of the complex $\omega$ under OBCs in both the $x$ and $y$ directions colored by the inverse participation ratio $\mathrm{IPR}(\psi_n)=\sum_i|\psi_n(\mathbf r_i)|^4/(\sum_i|\psi_n(\mathbf r_i)|^2)^2$ of the real-$\omega$ eigenstates $\psi_n$ by solving Eq.~\eqref{linear}. The larger the IPR is, the more localized the eigenstate is. The highly localized states are found in the gaps of $\omega=0$ and $-M^{-1}$. Fourfold degenerate gapped states with the largest IPR are present at $\omega=-M^{-1}$ and a set of eigenstates with large IPR form gapless boundary states around $\omega=0$. To reveal the feature of the latter, we plot in Fig. \ref{F1}(b) the spectrum in the strip geometry with OBCs along $x$ and PBCs along $y$. It shows that the boundary branches traverse the bulk gap and cross at $\omega=0$. Combining with the boundary distribution in Fig. \ref{F1}(c), this confirms the first-order topological nature of the set of gapless boundary states. In contrast, the states at $\omega=-M^{-1}$ are localized at the four corners [see Fig.~\ref{F1}(d)], identifying their second-order topological nature.

Figure \ref{F2}(a) shows the $\lambda$ spectrum of the auxiliary system colored by $\mathrm{IPR}$ under the OBC in both the $x$ and $y$ directions. Highly localized states intersect with $\lambda = 0$ just at $\omega_\star=-M^{-1}$ and around $\omega_\star=0$, corresponding exactly to the two types of boundary modes in Fig. \ref{F1}(a). These two $\omega_\star$ modes are well captured by $p_0(\omega)=0$, see the cyan dashed line in Fig. \ref{F2}(a). Therefore, the topological invariants defined in $P(k,0)$ and $P(k,-M^{-1})$ can characterize the boundary states in Fig.~\ref{F1}(c) and the corner states in Fig.~\ref{F1}(d), respectively. The $\lambda$ spectrum of $P(k,-M^{-1})$ in different $m$ is shown in Fig. \ref{F2}(b). $\mathcal{W}_2=1$ in Fig. \ref{F2}(c) succeeds in witnessing the presence of the second-order gapped corner states. With $m$ decreasing from the band closing point ${v\over 4M^2}-{2r\over M^2}+4w=-5v$ to another one ${v\over 4M^2}-{2r\over M^2}-4w=-9v$, the corner states disappear and the system becomes a first-order topological phase. Its nonzero topological invariant is changed to be $\mathcal{C}_2=\pm 1$. This phase is further verified by the spectrum under the strip geometry when $m=-6v$, where a pair of gapless states are present [see Fig. \ref{F2}(d)]. In the band gap of $\omega_\star=0$, the system hosts only the first-order topological phase described by ${\cal C}_1$. Figure \ref{F2}(e) shows the phase diagram of the zero mode in the $w$-$m$ space. The ones of the first- and second-order phase diagrams of the $-M^{-1}$ mode are presented in Figs. \ref{F2}(f) and \ref{F2}(g), respectively. The overlap of the regions with ${\cal C}_1\neq0$ and ${\cal W}_2=1$ yields a hybrid-order topological phase, with coexisting first-order boundary states at the band gap $\omega_\star=0$ and the second-order corner states at $\omega_\star=-M^{-1}$. 

\begin{figure}[ttp]
\begin{center}
\includegraphics[width=\columnwidth]{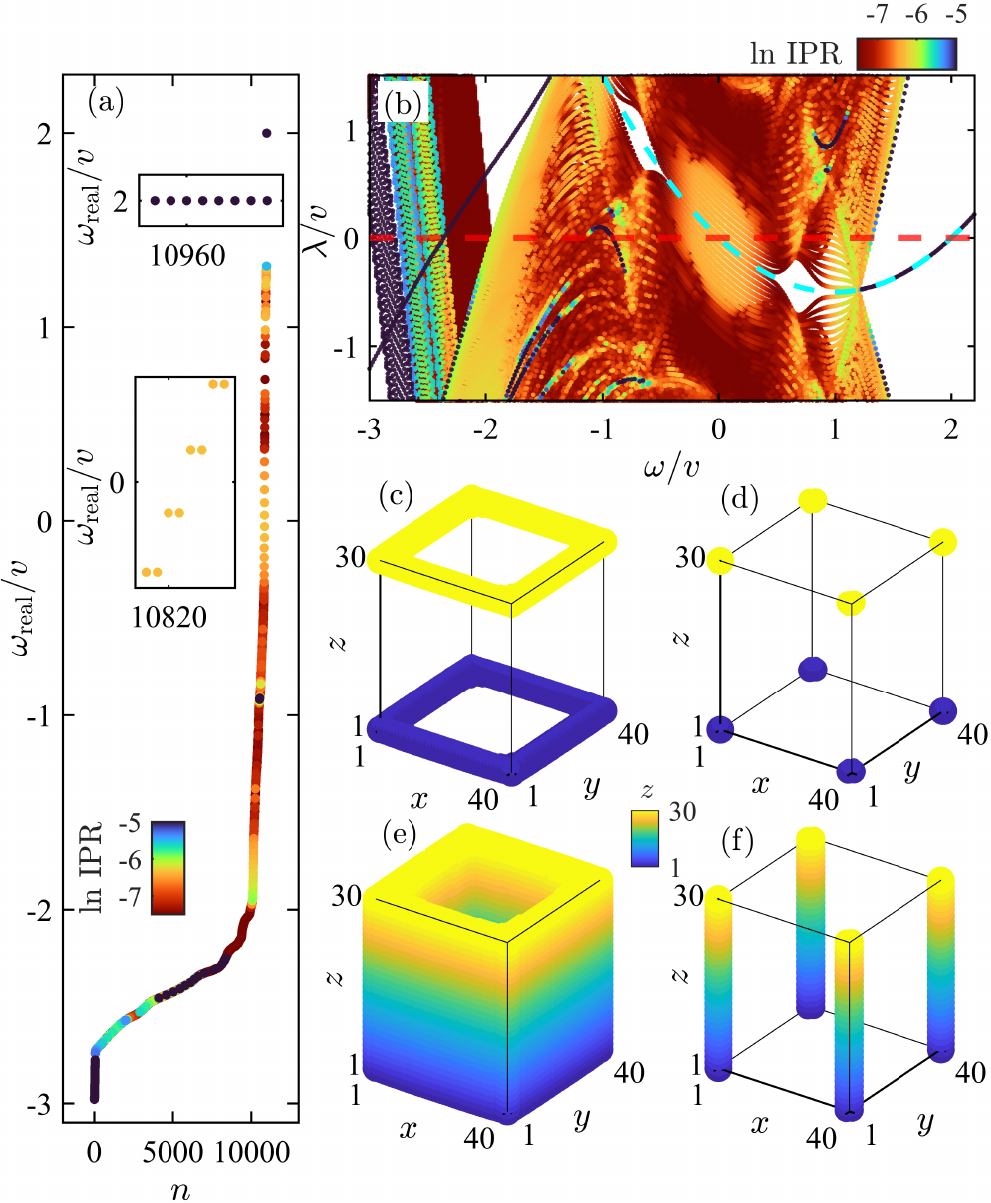}
\end{center}
\caption{Three-dimensional nonlinear hybrid-order topological phase. (a) Real nonlinear eigenvalue spectrum under full OBCs. The color determined by $\ln{\rm IPR}$. The insets show the localized states near $\omega=-M^{-1}$ and $\omega=0$. (b) Auxiliary spectrum of $P_{3D}(\mathbf{k},\omega)$. The red dashed line marks $\lambda=0$, the cyan dashed line shows $p_0(\omega)$. (c) and (d) is hinge state at $\omega=0$ and eight corner states at $\omega=-M^{-1}$. We take $t_1=1$ and $t_2=15$. For (e) and (f), we take $t_1=t_2=1$. The state at $\omega=0$ is a side-surface state at (e), while the state at $\omega=-M^{-1}$ is a hinge state at (f). The (c-f) are colored by $z$.}\label{F3}
\end{figure}
\section{3D hybrid-order nonlinear topological phases}
The generalization of our result to 3D systems can generate rich hybrid-order topological phases. We construct a 3D model by stacking the 2D nonlinear system along the $z$ axis through a SSH chain. In the basis $c_{\mathbf r}=(c_{\mathbf r,A,+},c_{\mathbf r,B,+},c_{\mathbf r,A,-},c_{\mathbf r,B,-})^T$, its Hamiltonian reads
\begin{equation}
    H_\text{3D}=H\tau_0+\sum_{\mathbf r}\{t_1c_{\mathbf r}^{\dagger}\sigma_0\tau_xc_{\mathbf r}+t_2[c_{\mathbf r+\hat z}^{\dagger}\sigma_0\tilde{\tau}_-c_{\mathbf r}+\text{h.c.}]\},\nonumber
\end{equation} where $\tilde{\tau}_-=(\tau_x-i\tau_y)/2$ and $t_{1/2}$ are the intracell and intercell hopping rates along the $z$ axis. The nonlinear part is ${\cal S}_\text{3D}(\omega)=\mathcal{S}(\omega)\tau_0$. Here, $H$ and $\mathcal{S}(\omega)$ are the same as those in the 2D case. The similar linearization as Eq.~\eqref{linear} turns the model into an eight-band system. Under the PBC, the pencil matrix is $P_\text{3D}(\mathbf k,\omega)=P({\bf k},\omega)\tau_0+\sigma_0[q(k_z)\tau_++q^*(k_z)\tau_-]$, where $q(k_z)=t_1+t_2e^{-ik_z}$ and $\tau_\pm=\big(\tau_x\pm i\tau_y\big)/2$. The separable structure of $P_{\rm 3D}({\bf k},\omega)$ between $k_{x/y}$ and $k_z$ makes its eigenstates factorized. The eigen equation of the 2D sliced subsystem is $P({\bf k},\omega)|\phi_s({\bf k},\omega)\rangle=s\rho({\bf k},\omega)|\phi_s({\bf k},\omega)\rangle$, with $s=\pm$ and $\rho({\bf k},\omega)=\sqrt{p_z^2+p_+p_-}$. The SSH block $h_z=q(k_z)\tau_++q^*(k_z)\tau_-$ has eigenvalues $\eta |q(k_z)|$, with $\eta=\pm$, and eigenstates $|u_\eta(k_z)\rangle$. Thus, the eigenvalues and eigenstates of $P_{\rm 3D}({\bf k},\omega)$ are $\lambda_{s,\eta}(\mathbf k,\omega)=p_0(\omega)+s\rho({\bf k},\omega)+\eta |q(k_z)|$ and $|\phi_s({\bf k},\omega)\rangle\otimes |u_\eta(k_z)\rangle$. The system follows the symmetries of 2D system multiplying the chiral system along the $z$ axis. This implies that the $xy$ topology of $P_{\rm 3D}({\bf k},\omega)$ is inherited from the 2D pencil matrix $P({\bf k},\omega)$, while its $z$-axis localization is controlled by the SSH chain. The $\omega$ spectrum of the nonlinear system is selected by the $\lambda=0$ line. The gap closings of the $\lambda$ spectrum at $p_0(\omega)=0$, giving $\omega_\star=0$ and $-M^{-1}$, can occur in two ways: either the 2D sliced subsystem and the SSH block close simultaneously, with the system parameters satisfying Eqs.~\eqref{dshl}-\eqref{ddsd} and $|t_1|=|t_2|$, or the two branches accidentally touch when $\rho({\bf k},\omega)=|q(k_z)|\neq0$. In the latter case, because both the 2D sliced subsystem and the SSH block remain gapped, this touching does not induce a topological phase transition.

When $|t_1|<|t_2|$, the system exhibits localization along the $z$ axis. Further if $P({\bf k},\omega_\star)$ has a nonzero Chern number, then the system also possesses 1D gapless boundary states in the $xy$ plane. Combining these two specific distributions, we obtain 1D hinge states in the 3D system, which is a second-order topological phase. If $P({\mathbf k},\omega_\star)$ has nonzero mirror-graded winding number, then the system also possesses gapped corner states in the $xy$ plane. Their combination results in corner states, which is a third-order topological phases. In the undimerized case $t_1=t_2$, the state along the $z$ axis becomes a metal state. Then $P({\bf k},\omega_\star)$ in a first-order topological phase of the 2D sliced subsystem supports the formation of the surface states in the 3D system after the $z$-axis stacking, signifying a 3D first-order topological phase. $P({\bf k},\omega_\star)$ in a second-order topological phase of the 2D sliced subsystem forms the hinge states in the 3D system, which is a 3D second-order topological phase. 

Choosing $t_1<t_2$, we show in Fig.~\ref{F3}(a) the real-$\omega$ spectrum colored by ${\rm IPR}$ under the OBC by solving Eq.~\eqref{linear}. The highly localized eightfold-degenerate gapped states are found at $\omega_\star=-M^{-1}$. A series of gapless localized states are present near $\omega_\star=0$. Figure~\ref{F3}(b) shows the $\lambda$ spectrum of the auxiliary system as a function of $\omega$. Its zero line correctly recovers the real-$\omega$ spectrum in Fig. \ref{F3}(a). The probability of the localized states near $\omega_\star=0$ in Fig. \ref{F3}(c) shows that they distribute on the hinges. It indicates the 3D system is in a second-order topological phase.  The probability distribution in Fig.~\ref{F3}(d) confirms the corner-mode nature of the gapped states at $\omega_\star=-M^{-1}$. This indicates that the 3D system exhibits a third-order topological phase. The above results verify the coexistence of the second- and third-order topological phases in our 3D system. On the other hand, choosing $t_1=t_2$, we realize the coexisting first- and second-order topological phases in the 3D system, see Figs.~\ref{F3}(e) and \ref{F3}(f). This numerical result matches our analytical expectation in the preceding section. Therefore, via well developed stacking engineering in layered structures, we can create rich hybrid-order topological phases in our nonlinear system.

\section{Discussion and conclusions}
Nonlinear-eigenvalue problems arise naturally in a broad range of wave systems. A standard oscillator model already gives a conventional quadratic eigenvalue problem. 
This makes quadratic eigenvalue dependence natural in mechanical, phononic, and acoustic systems \cite{10.1115/1.4026911,ZHOU2012041001,Chaigne2014,scjy-dbk7,10.1098/rspa.2025.0520,SHAABAN2026115891}. In dispersive optical and electromagnetic media, including waveguides, filters, and resonators, frequency-dependent material responses or effective boundary conditions make the resonant-mode equation nonlinear in the eigenfrequency \cite{Binkowski2019,Zolla:18,Joshi:26,PhysRevB.50.16835,PhysRevA.78.033834,8937095,8358010,PhysRevE.106.035304,LI2022108835,photonics10050523,PhysRevA.109.043518}. Topolectrical circuits provide another natural platform, where capacitive and inductive elements render the eigenvalue equation explicitly frequency-dependent \cite{Lee2018}. Therefore, our system can be conceptually derived from the design principles of these practical systems.

We have discovered hybrid-order topological phases in nonlinear-eigenvalue systems. A complete topological characterization is established through the introduction of an auxiliary system. In the 2D case, the hybrid-order topological phase is featured as the  simultaneous presence of the gapped corner states and gapless boundary states  within distinct band gaps of the nonlinear eigenvalue spectrum. Upon stacking the 2D system into 3D, we uncover a rich variety of hybrid-order topological phases. In the dimerized-stacked configuration, the corner states in the $xy$ plane become further localized at the two ends of the $z$ axis, manifesting as a 3D third-order topological phase, while the gapless boundary states in the $xy$ plane are localized at the hinges of the $z$ axis, manifesting as a 3D second-order topological phase. In the uniform-stack case, these two types of states transform into hinge states and side-surface states, respectively, indicating a coexisting first- and second-order topological phase. Our results open an avenue for exploring exotic topological phases through nonlinearity and stacking engineering. 

\section{Acknowledgments}
The work is supported by the National Natural Science Foundation of China (Grants No. 124B2090, No. 12275109, No. 92576202, and No. 12247101), the Quantum Science and Technology-National Science and Technology Major Project (Grant No. 2023ZD0300904), the Natural Science Foundation of Gansu Province (Grants No. 26RCKA011 and No. 25JRRA799), and the Fundamental Research Funds for the Central Universities (Grant No. lzujbky-2025-jdzx07).

\bibliography{BIB}
\end{document}